\documentclass[useAMS,usenatbib,usegraphicx]{mn2e}
\usepackage[utf8]{inputenc}
\usepackage[english]{babel}

\usepackage{amssymb} 
\usepackage{times} 

\usepackage{float} 
\usepackage{xcolor} 

\newcommand\ion[2]{#1$\;${\small\rmfamily{#2}}\relax}
\newcommand{\rfrag} {$R_{\rm frag}$}
\newcommand{\tfrag} {$t_{\rm frag}$}
\newcommand{\mfrag} {$M_{\rm frag}$}
\newcommand{\rs} {$R_{\rm s }{(0)}$}

\newcommand{\ms} {$\mathrm{M}_{\mathrm{\odot}}$}

\title[Star formation in the S233 region]{Star formation in the S233 region}

\author[Dmitry A. Ladeyschikov et al.]{D. A.
Ladeyschikov$^1$, A. M. Sobolev$^1$, S. Yu. Parfenov$^1$, S. A. Alexeeva$^2$, \newauthor J. H. Bieging$^3$\\
$^1$ Ural Federal University, 51 Lenin Str., Ekaterinburg 620051, Russia \\
$^2$ Institute of Astronomy of the Russian Academy of Sciences, 48 Pyatnitskaya Str., Moscow 119017, Russia \\
$^3$ Steward Observatory, University of Arizona, 933 North Cherry Avenue, Tucson, AZ 85721, USA \\}

\begin{document}

\date{Accepted \today, Received \today, in original form \today}
\pagerange{\pageref{firstpage}--\pageref{lastpage}} \pubyear{2014}
\maketitle
\label{firstpage}

\maketitle

\begin{abstract}
The main objective of this paper is to study the possibility of triggered star formation on the border of  the \ion{H}{II} region S233, which is formed by a B-star. Using high-resolution spectra we determine the spectral class of the ionizing star as B0.5~V and the radial velocity of the star to be $-17.5\pm1.4$ km~s$^{-1}$. This value is consistent with the velocity of gas in a wide field across the S233 region, suggesting that the ionizing star was formed from a parent cloud belonging to the S233 region. By studying spatial-kinematic structure of the molecular cloud in the S233 region, we detected an isolated clump of gas producing CO emission red-shifted relative to the parent cloud. In the UKIDSS and WISE images, the clump of gas coincides with the infrared source containing a compact object and bright-rimmed structure. The bright-rimmed structure is perpendicular to the direction of the ionizing star. The compact source coincides in position with IRAS source 05351+3549. All these features indicate a possibility of triggering formation of a next-generation star in the S233 region. Within the framework of a theoretical one-dimensional model we conclude that  the `collect-and-collapse’ process is not likely to take place in the S233 region. The presence of the bright-rimmed structure and the compact infrared source suggest that the `collapse of the pre-existing clump’ process is taking place.
\end{abstract}


\section{Introduction} \label{sec:intro}

A considerable number of stars in our and other galaxies are formed as a result of star formation triggered by the expansion of \ion{H}{II} regions. Such expanding \ion{H}{II} regions have a great variety. They can appear as a result of rare energetic events like a supernova explosion producing huge superbubbles \citep{Kang}, rare energetic objects like WR stars or clusters of massive stars \citep{Oey05,Schneider10}. These energetic objects attract the majority of effort in the field of triggered star formation. At the same time star formation can be efficient on the borders of the \ion{H}{II} regions formed by almost isolated massive stars \citep{Kirsanova08, Snider09, Zavagno10}.

Rather frequently triggered star formation is realized as a collapse of a pre-existing clump. In this process contraction of a pre-existing molecular clumps is initiated by the photoionization and shock fronts propagating away from the \ion{H}{II} region. Bright rims are expected to appear at the sides of the clumps which face the ionizing star (similar to those observed e.g. by \citet{Thompson04}). Clear examples of the star formation due to compression of pre-existing dense condensations are found in IC~1396, IC~1805, SFO~79 star-forming regions \citep{Weikard96,Heyer96b,Urquhart04}.

Another type of triggered star formation is the `collect-and-collapse’ process, first proposed by \citet{Elmegreen77}. In this process massive OB-stars create an \ion{H}{II} region expanding into the ambient molecular cloud due to the presence of a pressure difference between the molecular and ionized gas. This expansion compresses ambient molecular material, consecutive fragmentation of which creates molecular clumps and filaments. These clumps are progenitors for the next generations of stars.  The important aspect of this scenario is that there is a delay between the expansion of the \ion{H}{II} region and the onset of triggered star formation \citep{Elmegreen98}. This is different from the case of star formation by collapse of pre-existing clumps where the formation of the star starts almost immediately. Clear examples of star formation by the `collect-and-collapse’ process  in our Galaxy can be found in S104 and S212 \citep{Deharveng03,Deharveng08}. 

\citet{Whitworth94} investigated properties of the `collect-and-collapse’ process of star formation with a one-dimensional model of the shocked shell driven by expansion of the \ion{H}{II} region into a uniform medium. They concluded that resulting fragments are likely to have high mass ($\gtrsim 7$~\ms). It was found that gravitational fragmentation of the shell takes place when the column density of hydrogen in the shell reaches $\approx6\times10^{21}$~cm$^{-2}$. \citet{Dale09} use the thee-dimensional smooth particle hydrodynamic (SPH) model to test the applicability of the thin-shell approximation used in \citet{Whitworth94} for more realistic astrophysical situation. They concluded that this approximation should be used with some caution because of the problem with the boundary conditions which greatly influence the fragmentation of the shell.

In this paper, we consider the border of an \ion{H}{II} region around an early-type star. It will be shown that  there are signs of the considerable influence of this ionizing star on the ambient molecular cloud. This influence possibly results in triggering the formation of the next-generation star(s) on the border of the S233 \ion{H}{II} region.

The S233 region is a small ($\simeq 2\arcmin$ in diameter) optical emission nebula formed by the ionizing star USNO-A2 1200-03588518 ($\alpha_{J2000}=05^{\rmn{h}}38^{\rmn{m}}31\fs 5, \delta_{J2000}= +35\degr 51\arcmin 19\arcsec$) with visual magnitude $V=11.7^m$. The S233 region is a part of S231-235 star formation complex. This complex is located in the Perseus spiral arm and contains four developed \ion{H}{II} regions: S231, S232, S233, and S235 \citep{Heyer96a}. Published distance estimates to the objects in this complex vary from 1.6 to 2.3 kpc \citep{Reiputh}. 
The S233 region is mentioned in the review of \citet{Reiputh}, which states that the S233 \ion{H}{II} region is excited by a B1.5~II star \citep{Hunter}. In the present study we conducted a spectroscopic analysis of the ionizing star using high-resolution spectroscopic data and have shown that it is a main sequence star of the B0.5 spectral type with mass of $13\pm1$~\ms{}.

\citet{Wouterloot} studied CO(1-0) molecular line emission toward the IRAS sources beyond the solar circle. They distinguished two separate components of the CO(1-0) line toward IRAS 05351+3549 situated within the optical S233 emission. Our newly obtained data enable us to investigate the spatial structure of these CO components. The main gas component of CO emission ($12.1 \pm 0.4~\rm{K}$ at $-18.2 \pm 0.4~\rm{km~s^{-1}}$) has a large-scale spatial extension, but the second component ($2.6 \pm 0.4~\rm{K}$ at $-12.9 \pm 0.3~\rm{km~s^{-1}}$) is a rather compact clump of gas situated close to the infrared source IRAS 05351+3549.

\section{Observations and data reduction}

\subsection{The $^{12}$ CO and $^{13}$ CO Molecular Gas} \label{sec:smt_data}

For our study of molecular gas in the vicinity of the S233 region, we used information on emission in four spectral lines of the CO molecule: $^{12}$CO(1-0), $^{12}$CO(2-1) and the lines of isotopologue $^{13}$CO(1-0) and $^{13}$CO(2-1). These lines are known to be an efficient probe of morphology and physical parameters of molecular gas.

The maps of the $^{12}$CO(2-1) and $^{13}$CO(2-1) spectral lines were obtained with the 10-meter Heinrich Hertz Submillimeter Telescope (SMT), a facility of the Arizona Radio Observatory. The maps include a $70\arcmin \times 50\arcmin$ region centered at $\rmn{RA}(2000)=5^{\rmn{h}} 40^{\rmn{m}}$, $\rmn{Dec.}(2000) = 35\degr 50\arcmin$ which covers more than $20\arcmin$ around the vicinity of the S233 region. Observations were obtained in January 2010. The whole map consists of subfields with $10\arcmin \times 10\arcmin$ size. Each subfield was mapped using the On-The-Fly mode of the SMT telescope. The beam size was 32\arcsec{} for the $^{12}$CO(2-1) and 33.5\arcsec{} for the $^{13}$CO(2-1). Observed frequencies were 230.53800~GHz for the $^{12}$CO(2-1) and 220.39868~GHz for the $^{13}$CO(2-1) lines, respectively. The main beam efficiency was 0.81 for the $^{12}$CO line and 0.78 for the $^{13}$CO line. The observations were made with a 10\arcsec~angular spacing, which was at least three times smaller than the beam size. This allows avoiding sudden changes in intensity from pixel to pixel and positively affects the accuracy of determination of physical parameters in the molecular cloud. The spectral resolution was 0.325 $\rmn{km~s}^{-1}$ for the $^{12}$CO(2-1) line and 0.340 $\rmn{km~s}^{-1}$ for the $^{13}$CO(2-1) line. Achieved RMS noise levels were 0.23~K and 0.21~K for the $^{12}$CO(2-1) and $^{13}$CO(2-1) lines, respectively.

The $^{12}$CO(1-0) and $^{13}$CO(1-0) spectral lines were mapped using the 13.7-meter FCRAO radio telescope with the 32-pixel SEQUOIA focal plane array receiver. The S231-S235 complex was mapped in January 2000. The map of each isotope line includes a $150\arcmin \times 150\arcmin$ region based on the galactic coordinate system, centered at $l=173.25\degr$, $b=2.75\degr$ ($\rmn{RA}=5^{\rmn{h}} 40^{\rmn{m}}$, $\rmn{Dec.}= 36\degr 07\arcmin$). The main beam sizes were 45\arcsec~and 47\arcsec~for the $^{12}$CO(1-0) and $^{13}$CO(1-0) lines, respectively. Observed frequencies were 115.27120~GHz and 110.20135~GHz for the $^{12}$CO(1-0) and $^{13}$CO(1-0) lines, respectively. The pixel spacing was 22.5\arcsec{} and the velocity channel spacings were 0.127 $\rmn{km~s}^{-1}$ and 0.133 $\rmn{km~s}^{-1}$ for the $^{12}$CO(1-0) and $^{13}$CO(1-0) lines, respectively. Achieved RMS noise levels were 1.1~K and 0.63~K for the $^{12}$CO(1-0) and $^{13}$CO(1-0) lines, respectively.

The main beam efficiencies for the \mbox{$^{12}$CO(1-0)} and \mbox{$^{13}$CO(1-0)} lines were 0.45 and 0.48, respectively. These efficiencies applied if the source of emission was approximately the same size as the main beam. However, if the emission is widely distributed on scales of ~$0.5\degr$ or more, it fills the error beam of the FCRAO telescope, and in such a case one need to apply scattering efficiency of 0.7 for both lines. Given the large range of angular sizes of the structures in S233 complex neither main beam nor scattering efficiencies will give good calibration. To calibrate the FCRAO CO(1-0) data we used a deconvolution method described in \citep{Pineda} to remove a $0.5\degr$ error beam component and then we used main beam efficiencies. The deconvolution method involves a division of Fourier transform of source data by Fourier transform of the error beam. Correction for the error beam improves the calibration of the line intensities by 10-30 percent.

We used the \textsc{miriad} software package \citep{Sault} for manipulating data cubes and the \textsc{karma} package \citep{Gooch} for visualization. In this paper, we have extracted a $10\arcmin \times 10\arcmin$ region centered at the position of the ionizing star in the S233 region with coordinates $\rmn{RA}(2000)=5^{\rmn{h}} 38^{\rmn{m}} 31\fs5$, $\rmn{Dec.}(2000) = 35\degr 51\arcmin 19\arcsec$. Because the data for different transitions was obtained with different receivers, we have applied some corrections to the original data for comparing line profiles observed with different telescopes.

 These corrections include the convolution of all data cubes to the same beam size of FWHM$=47\arcsec{}$. For the $^{13}$CO (1-0) line this is the original beam size, thus no changes were made to this data cube, but for the $^{12}$CO (2-1) and $^{13}$CO (2-1) lines the original beam size is 32\arcsec{} and 33.5\arcsec, respectively. Thus to achieve 47\arcsec{} beam size we convolved these line maps with $\sqrt{47^{2}-32^{2}}=34.4\arcsec$ and $\sqrt{47^{2}-33.5^{2}}=32.9\arcsec$ Gaussians, respectively.  For investigating the spatial-kinematic structure of molecular cloud the corrections are not necessary.

\subsection{Echelle-spectrum of the ionizing star}

The observations were obtained in January 2012 with the Echelle-spectrograph mounted on the Special Astrophysical Observatory (SAO) 6-m telescope \citep{Panchuk}. The detector of the Echelle-spectrograph was a large size $4608\times 2048$-pixel CCD array with an image slicer. We used a setup yielding a resolving power of R = 50000, with a spectral coverage of 4000-7000 \AA. A hollow cathode Th-Ar lamp was used for the wavelength calibration.

We have obtained 6 spectra with a 45 minutes exposure time for each spectrum. Initial inspections showed no significant velocity variation in this sample. Thus, all spectra were combined to improve the signal-to-noise ratio (SNR) and a
master spectrum was created for subsequent analyses. The initial steps of the data reduction process (removal of cosmic ray features, background subtraction, and spectral order extraction) were done under the \textsc{echelle} context of \textsc{midas} software package \citep{Yushkin}, while the final steps (i.e., continuum normalization, radial velocity and equivalent widths (EW) measurements) were performed using the package \textsc{dech20} \citep{Galazutdinov}. The SNR of the reduced spectra depends on wavelength, and is about 60 and 30 in H$\alpha$ (6563 \AA) and H$\delta$ (4861 \AA) regions, respectively.

\begin{figure}
\includegraphics[width=84mm]{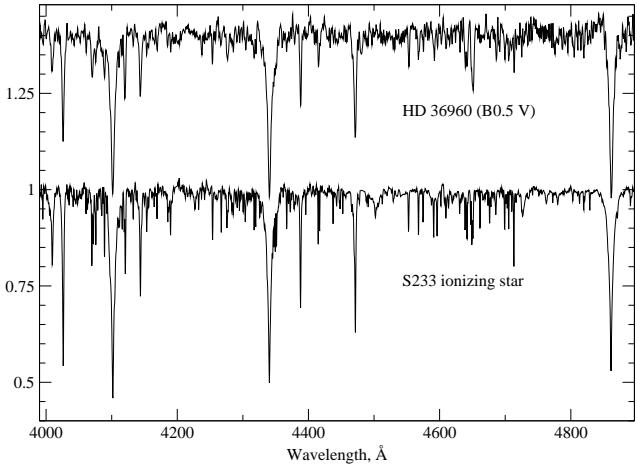}
\caption{Smoothed and rebinned to R=4000 spectrum of S233 ionizing star (lower spectrum) and spectrum of HD 36960 (B0.5 V) from the digital atlas of \citet{Walborn}.}
\label{fig:S233_spectra}
\end{figure}

\subsection{Spectral Analysis}\label{sec:spec_analys}

As noted in Section~\ref{sec:intro} the S233 ionizing star was classified as B1.5~II by \citep{Hunter}. We classified this star as B0.5~V by comparing our smoothed and degraded to R = 4000 spectrum with the corresponding part of the digital atlas of \citet{Walborn} (see Figure~\ref{fig:S233_spectra}). Degradation was performed using the GNU Data Language routine \textsc{`rebin.pro'}. The main criteria to estimate the spectral type of the ionizing star were approximate equality of the strengths of the Si III 4552 and Si IV 4089 lines and the presence of the very weak He II 4686 line \citep{Walborn}. The ratio of Si III 4552 and He I 4837 lines was the primary criterion to estimate the luminosity class. It seems that the modest disagreement in the spectral classification is likely due to differences in the resolution and perhaps SNR, together with some degree of subjective judgment in the comparison by eye with standard spectra.

To confirm the spectral classification we also performed the spectroscopic analysis using stellar atmosphere models from the \textsc{bstar2006} grid with the solar metallicity and microturbulent velocity of $v_{\rm{turb}}=2$ km~s$^ {-1} $ \citep{Lanz}. The steps of the effective temperature $T_{\mathrm{eff}}$ and surface gravity (log$\,\textsl{g}$ ) in the grid are 1000~K and 0.25~dex, respectively.

At the first step of the spectral analysis the projected rotational velocity $v\sin i$ was obtained by comparing observed line profiles of helium and metal lines with synthetic ones. Synthetic profiles were computed with the values of $T_{\mathrm{eff}}$ = 29000~K and log$\,\textsl{g}=4.0$~dex which correspond to a B0.5~V star \citep{Simon-Diaz10}. The comparison was performed by eye using synthetic spectra calculated by the \textsc{rotin3} program (provided with \textsc{synspec} \citep{Hubeny}) for three $v\sin i$ values: 10, 20 and 30 km~s$^{-1}$. The relatively large step of 10 km~s$^{-1}$ allows to easily determine how well the observed spectrum is fitted by a synthetic spectrum. Spectra with $v\sin i < 10$~km~s$^{-1}$ and $v\sin i > 30$~km~s$^{-1}$ gave a very poor fit. The resulting value of $v\sin i$ and its standard deviation are 13~km~s$^{-1}$ and 5~km~s$^{-1}$, respectively.

At the second step of the spectral analysis the observed spectrum was compared with a number of synthetic spectra computed using \textsc{bstar2006} grid for different values of $T_{\mathrm{eff}}$ and log$\,\textsl{g}$. We compared observed profiles of Si III $\lambda\lambda$4552, 4567, 4574, 4716, 4813, 4819, 4829, 5740 and Si IV $\lambda\lambda$4089, 4116, 4631 and cores and wings of Balmer lines (H$\alpha$, H$\beta$, H$\delta$, H$\gamma$). To characterize how well the observed profiles are fitted by model spectra we used the quantity $\chi^2$

\begin{center}
\begin{equation}
\chi^2=\frac{1}{n_\mathrm{lines}}\,\sum_{\mathrm{i=1}}^{n_\mathrm{lines}} \frac{w_\mathrm{i}}{n_{\nu}}\,\sum^{n_{\nu}}_{\mathrm{j}=1} \left(\frac {y_\mathrm{j}^\mathrm{i}-y_\mathrm{j\,obs}^\mathrm{i}}{\sigma^\mathrm{i}}\right) ^{2}
\label{Eq:criterium}
\end{equation}
\end{center}

\noindent where $n_{\nu}$ is the number of wavelength points in the spectral line $i$, $n_\mathrm{lines}$ is the number of spectral lines used in the analysis, $y_\mathrm{j\,obs}^\mathrm{i}$ is the observed flux in the $j$-th point of the spectral line $i$, $y_\mathrm{j}^\mathrm{i}$ is the synthetic flux, $w_\mathrm{i}$ is a weight which corresponds to the spectral line $i$, and $\sigma^\mathrm{i}=\mathrm{(SNR)}^{-1}$ accounts for the signal-to-noise ratio of the spectral line $i$. For all lines except Si IV $\lambda$4089 a value of weights equals to 1.0 were used. For the Si IV $\lambda$4089 line we used a weight of 0.5 because it was blended with the OII line.

\begin{figure}
\includegraphics[width=84mm]{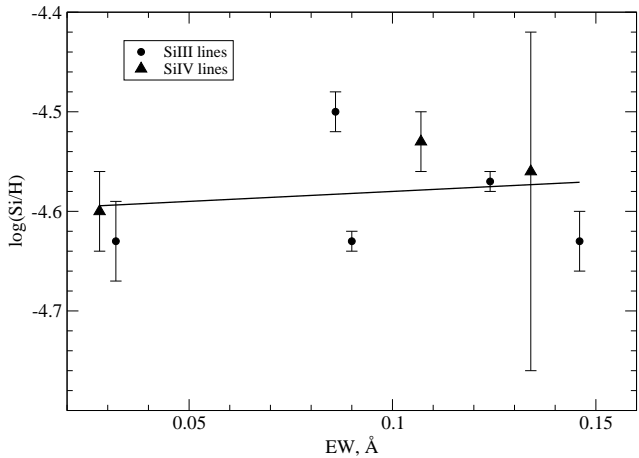}
\caption{Silicon abundance vs. EW diagnostic diagram obtained for $v_{\rm{turb}}=6$ km~s$^{-1}$. The solid line represents the linear regression fit. Uncertainties in the individual line abundances (propagated from the standard deviations in the measured EWs) are indicated as error bars. Because Si IV $\lambda$4089 line was blended with the OII line, second to last point has large uncertainty. }
\label{Siabund}
\end{figure}

The minimization of $\chi^2$ provides the maximum likelihood estimate of the model parameters, $T_{\rm{eff}}$ and log$\,\textsl{g}$. In our case the model with $T_{\rm{eff}}=28000$~K and log$\,\textsl{g}=4.0$ exhibit the minimum of $\chi^2$. These model parameters were used to evaluate silicon and helium abundances based on the curve-of-growth method (see, for example, \citet{Kilian92,Simon-Diaz10}). In this method the abundance and $v_{\rm{turb}}$ for a given $T_{\rm{eff}}$ and log$\,\textsl{g}$ are varied while the same abundance is not obtained for all lines. We used a grid of synthetic spectra for various abundances and $v_{\rm{turb}}$ values that were computed with the code \textsc{synspec48} \citep{Hubeny}. The number densities of Si in the grid are $\rm{log(Si/H)} = -4.75, -4.57, -4.45, -4.32, -4.15$ dex where $\rm{log(Si/H)} = -4.45$ is the solar abundance \citep{Grevesse}. The He/H values are varied from 0.05 to 0.15 with 0.025 step. Using a linear interpolation on the grid with $v_{\rm{turb}}=6$ km~s$^{-1}$ we have obtained the log(Si/H)-EW diagram that is shown in Figure~\ref{Siabund}. The log(Si/H)-EW dependence was approximated by the straight line that is shown as solid line in Figure~\ref{Siabund}. The linear approximation of log(Si/H)-EW is given by $\mathrm{log(Si/H)}=-4.60\pm0.05+(0.2\pm0.5)\mathrm{EW}$, where after a `$\pm$'~sign standard errors are given. It should be noted that Si III $\lambda\lambda$4813, 4819, 4829 lines were excluded from the Si abundance analysis because this triplet tends to give different results. This discrepancy could be related to the problems of a Si model atom (see e.g. \citet{Becker90}). Using the similar diagram, we also obtained the helium abundance $\mathrm{He/H}=0.10\pm0.01$.

\begin{figure}
\includegraphics[width=84mm]{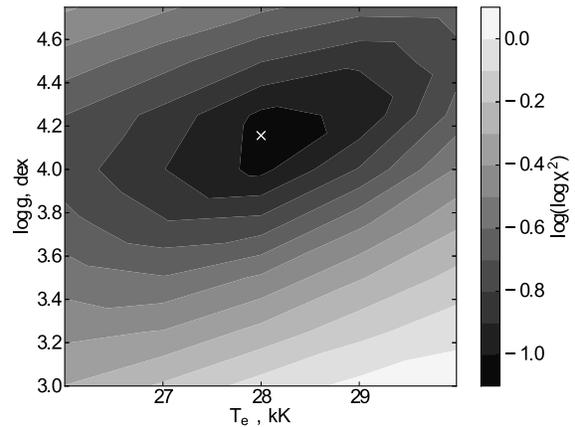}
\caption{$\chi^2$ difference between the observed spectrum and spectra computed with $v_{\rm{turb}}=6$ km~s$^{-1}$ and $\rm{log(Si/H)}=-4.60$. Abundances of other chemical elements are solar. The $\chi^2$ values are quantified by shades of grey. The cross marks the position of the model with minimum value of $\chi^2$.}
\label{khi_distr}
\end{figure}

At the final step of spectral analysis we obtained a $\chi^2$ distribution for $v_{\rm{turb}}=6$ km~s$^{-1}$ and $\rm{log(Si/H)}=-4.60$. This distribution is shown in Figure~\ref{khi_distr}. According to this distribution, we selected all models with a $\chi^2$ below the critical value. The critical value is a minimal $\chi^2$ value for models on the edge of the grid. The resulting values of $T_{\rm{eff}}=28000\pm1100$ K and log$\,\textsl{g}=4.15\pm0.27$ dex are obtained by averaging selected models, weighted by $\exp{(-0.5\chi^2)}$, and their standard deviations.

A comparison of our derived parameters with the calibration of \citet{Crowther} enables us to suggest that the S233 ionizing star is a B0.5~V star. The He abundance also indicates that most likely this star is still on the main sequence. We estimated the mass of the S233 ionizing star to be $\mathrm{M/M_{\bigodot}}=13\pm1$ by interpolating the stellar evolutionary tracks of \citep{Claret} in the log$\,\textsl{g}$-$T_{\rm{eff}}$ plane (see, e.g., \citet{Kilian92}) and using the Monte-Carlo error propagation. An estimate of the age is much more uncertain with the standard deviation equal to the value itself: $(3\pm3)10^6$ yr. Nonetheless, one may say that the star is rather young and spent about $0.2\pm0.2$ of its lifetime on the main sequence.

To estimate the distance to the ionizing star we used the K-band magnitude $K=9.63$ obtained from 2MASS, absolute magnitude $M_K=-2.512$ \citep{Bertelli} and interstellar absorption $A_K=A_V/8.8=0.35$ according to the reddening law of \citet{Cardelli} and $A_V=3.1$ estimated by \citet{Dobashi}. The estimated distance $2.3\pm 0.4$ kpc is in agreement with published distances to the S231-235 complex. For example, \citet{Chan} estimated the distance to the S233 region as $2.3\pm 0.7$ kpc.

The range of the effective temperature of the ionizing star ($T_{\mathrm{eff}}=28000\pm1000$~K) limits the range of the number of ionizing photons $\log N_{\mathrm{c}}$, to between 47.0 to 47.34. This range is obtained by linear extrapolation of values on $T_{\mathrm{eff}}$ in Table 4 of \citet{Martins05}. \citet{Hunter} give two estimates of the number of ionizing photons: $\log N_{\mathrm{c}}=47.45$ and $\log N_{\mathrm{c}}=47.59$, the values are estimated from radio continuum and H$\alpha$ observations, respectively. However, we note that they adopted a distance to S233 of 4.7 kpc. If we adopt the distance of 2.3~kpc, then the values of the number of ionizing photons should be scaled by $(4.7/2.3)^2=4.176$. This gives $\log N_{\mathrm{c}}=46.82$ and $\log N_{\mathrm{c}}=46.97$. The value $\log N_{\mathrm{c}}=46.97$ is almost equal to the lower estimate of the number of ionizing photons with the stellar parameters obtained by us.  This indicates that the ionizing star has a sufficient Lyman continuum flux to account for the observed H$\alpha$ intensity and the radio flux density of the S233~\ion{H}{II} region. Most likely estimation of the number of ionizing photons for the S233 \ion{H}{II} region is $\log N_{\mathrm{c}}=47.0$.

\subsection{Radial velocity measurements} \label{sec:rad_vel}

The radial velocities were measured for 150 absorption lines using \textsc{dech20}. The radial velocities were obtained interactively with a mirroring method \citep{Parimucha}, i.e by means of the best match of the line profile with its mirror. Rest wavelengths for each line were taken from VALD \citep{Piskunov}. The derived radial velocity of the S233 ionizing star is $-17.5\pm1.4$ km~s$^{-1}$. This velocity is consistent with the velocity of the surrounding molecular cloud (from $-$15 to $-$20 km~s$^{-1}$), thus the physical association with this molecular cloud may be assumed.

There are also the nebular lines in the Echelle-spectrum of the ionizing star. Using [SII] $\lambda\lambda$6716, 6731 and [NII] $\lambda\lambda$6548, 6583 lines we estimated the radial velocity $-10.7\pm1.7$ km~s$^{-1}$. Because the nebular lines are used, the derived velocity represents the ionized gas in the S233 region.

Analyzing the velocities of the star, ionized and molecular gas, we conclude that the ionizing star is moving together with the parent molecular cloud, but the ionization front is moving slower in the direction to the observer with a velocity of $6.8\pm2.2$~km~s$^{-1}$ with respect to the ionizing star. This value indicates the expansion velocity of the S233 \ion{H}{II} region.

\section{Results}

\subsection{Nebulosity in the S233 region} \label{sec:nebl}

\subsubsection{Optical images} \label{sec:optical}

\begin{figure*}
\begin{minipage}{140mm}
\includegraphics[width=135mm]{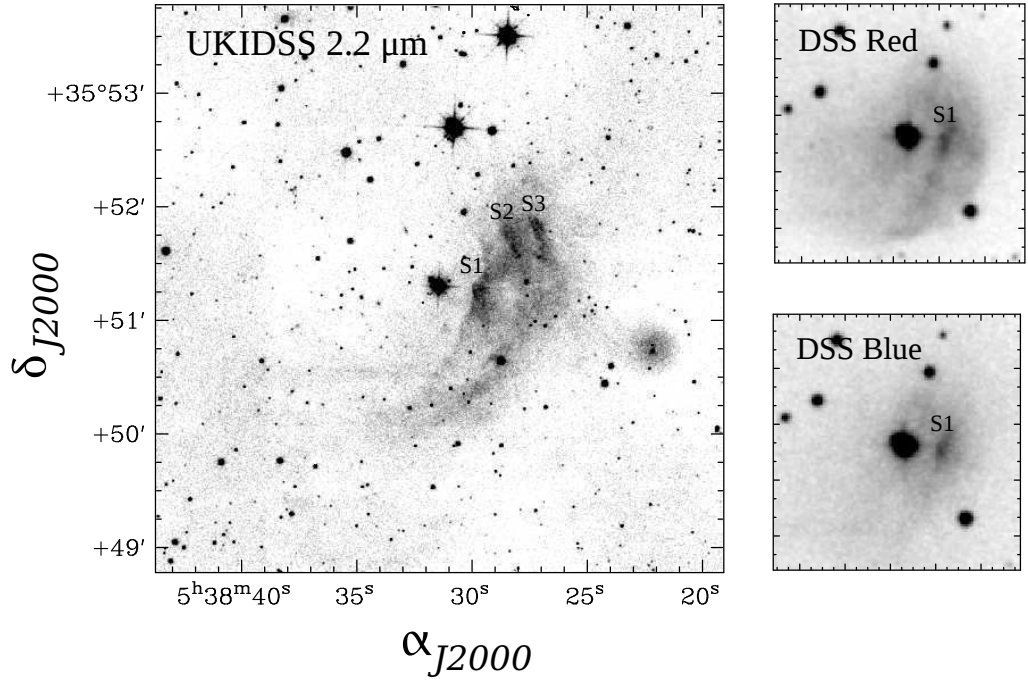}
\caption{UKIDSS near-IR (K-band) and DSS optical (Red and Blue) images of the S233 region. The center of the image coincides with the ionizing star at $\rmn{RA}(2000)=5^{\rmn{h}} 38^{\rmn{m}} 31\fs5$, $\rmn{Dec.}(2000) = 35\degr 51\arcmin 19\arcsec$. The image size is $5\arcmin\times 5\arcmin$. The condensations S1, S2 and S3 discussed in the text (see Section \ref{sec:nebl})  are marked with labels. }
\label{ukidss}
\end{minipage}
\end{figure*}

The S233 region is seen as an optical \ion{H}{II} region. In the DSS Red image the optical nebula exhibits almost a spherical morphology (see Figure~\ref{ukidss}) which is typically found in \ion{H}{II} regions. The DSS Red image traces primarily ionized gas. The ionized region does not have a clear edge at the north-east while there is a clear edge seen in the south-west side.

The DSS Red image also shows condensation at the west of the central ionizing star. This condensation is also seen in near-IR and mid-IR images (see Section~\ref{mid_IR} and \ref{near_IR}). The condensation is also detectable in the DSS Blue image, which cannot be explained as emission directly of the interstellar dust. If we assume the emission is the reflected light of the central ionizing star, the condensation detected in the DSS Blue image supports that the region is physically associated with the ionizing star.

\begin{figure*}
\begin{minipage}{140mm}
\includegraphics[width=135mm]{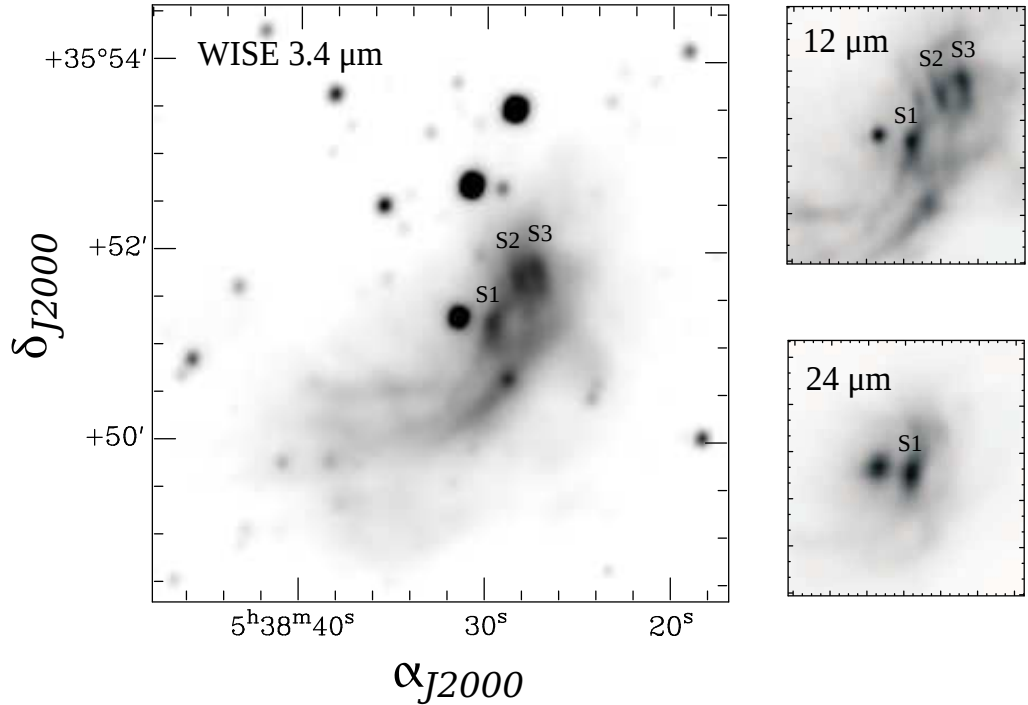}
\caption{WISE mid-IR images of the S233 region. The center of the left image coincides with the S233 ionizing star at position $\rmn{RA}(2000)=5^{\rmn{h}} 38^{\rmn{m}} 31\fs5$, $\rmn{Dec.}(2000) = 35\degr 51\arcmin 19\arcsec$. The image size is $5\arcmin\times 5\arcmin$. The left image shows distribution of 3.4 \micron{} emission, the two small images at the right shows the distribution of 12 \micron{} and 24 \micron{} emission. The resolution of 12 and 24 \micron{} images was improved with the HiRes technique \citep{Masci}. The condensations S1, S2 and S3 discussed in the text (see Section \ref{sec:nebl}) are marked with labels.}
\label{wise}
\end{minipage}
\end{figure*}

\subsubsection{WISE mid-IR images} \label{mid_IR}

Figure \ref{wise} shows the mid-IR images of the S233 region. These images were obtained from the all-sky data release \citep{Wright} of the Wide-field Infrared Survey Explorer (WISE) at 4.6, 12 and 24 \micron. The images reveal the existence of diffuse nebulosity to the south-west of the ionizing star. The main contribution to the emission at 12 \micron{} comes from polycyclic aromatic hydrocarbons (PAHs) in a photodissociation region (PDR). Usually a PDR appears as a relatively thin region of mid-IR emission along the periphery of an \ion{H}{II} region, where PAH is excited enough to glow, but not to be destroyed by UV-radiation

In our case we do not see a sharp boundary of the PAH emission of the S233 \ion{H}{II} region. This could be due to projection effects together with the non-uniform distribution of the ambient material. If we overlay the DSS Red image with the WISE 12 \micron{} image (see Figures~\ref{wise} and~\ref{ukidss}), we find that the optical emission of the ionized gas overlaps with the PAH emission recognized in the 12 \micron{} image. In the case of uniform distribution of material around a spherically expanding \ion{H}{II} region, the bright PAH emission should surround the ionized gas. In the present case the PAH emission overlaps with the optical emission of the ionized gas, which  could be due to projection effects.  This comparison suggests that the \ion{H}{II} region interacts with the ambient material on the far side from the observer, so the region of the PAH emission is relatively extended. Therefore, the conclusion is that the ambient material is denser in the south-west region: this situation is consistent with the cometary-like structure of the ionized hydrogen region, seen in red optical images.

The nebulosity contains a number of condensations, which are labeled on the Figure~\ref{wise} as S1, S2 and S3. The bright point source below S1 is likely a foreground star as it coincides with a star seen in the optical image, identified as  USNO-A2 1200-03587620 and 2MASS source 05382867+3550388 with a magnitude 12.5 in the K band.

In Figure~\ref{wise}, we see the emission of the ionizing star at 24 \micron. In order to check whether this emission is spatially separated from the nearest infrared peak S1, we obtained images with a higher spatial resolution using the HiRes technique \citep{Masci}. The resulting HiRes image of the S233 region shows a clear separation between the S1 condensation and the ionizing star emission. The S1 condensation is also seen in the DSS Blue band image as it reflects the light of the ionizing star. From this fact we can conclude that S1 is physically associated with the ionizing star in the S233 region (see Section \ref{sec:optical})

\subsubsection{UKIDSS near-IR images} \label{near_IR}

The UKIDSS survey \citep{Lawrence} is the next generation near-infrared sky survey, the successor to 2MASS. The survey instrument is \hbox{WFCAM} \citep{Casali} mounted on the UK Infrared Telescope (UKIRT) in Hawaii. The high resolution of \hbox{WFCAM} (0.4\arcsec) gives a high quality images of the particular regions of the sky. The S233 region lies in the area covered by the Galactic Plane Survey (GPS) of UKIDSS.

Figure~\ref{ukidss} shows the image in the \hbox{WFCAM} K band along with the optical images. It is clearly seen that in the S233 region there are well-pronounced structures having forms of bright rims elongated perpendicular to the ionizing star. Within these rim structures the condensations S1, S2 and S3 are pronounced. The fine structure of these condensations is clearly seen in this image. The S1 condensation has a  brightness gradient, revealing the brightest part on the ionizing star side and the darker part in the opposite side. A number of molecular clouds associated with similar structures and condensations were studied by \citet{Thompson04}.

\subsubsection{IRAS far-IR data} \label{sec_IRAS}

Finally, the IRAS source 05351+3549 is located in the vicinity of the S233 region. The angular separation between the IRAS source 05351+3549 and the S1 condensation, discussed in Section~\ref{mid_IR}, is 17\arcsec. Taking into account the uncertainty of the IRAS source position (major radius of the error ellipse is 31\arcsec), we assume that the IRAS source 05351+3549 is associated with the S1 peak seen in the mid-IR images of WISE and near-IR images of UKIDSS. This IRAS source has the following flux densities: $F_{12} = 1.09~\mathrm{Jy}$, $F_{25} = 11.5~\mathrm{Jy}$, $F_{60} = 184~\mathrm{Jy}$, $F_{100} = 414~\mathrm{Jy}$. It is notable that the flux density increases as the wavelength increases.  The colours (according to definition of \citet{Walker88}) of IRAS 05351+3549 are $[12]-[25]=4.12$, $[25]-[60]=4.89$. These colours correspond to those of a Young Stellar Object in the colour-colour diagram given by \citet{Straizys07}.

One may notice that IRAS 05351+3549 is marked an extended source in the IRAS catalog.  This extension is probably  due to the association of another infrared source with the S1 clump. Moreover, the central ionizing star, which is located close to the S1 peak ($d\simeq 24\arcsec$), emits radiation at mid-IR wavelength and falls into the ellipse of IRAS source position error. The angular resolution of the IRAS image is not sufficient to resolve the S1 peak and the ionizing star and to construct their reasonable spectral energy distributions. But mid-IR images from WISE show that these peaks can be distinguished (see Figure~\ref{wise}).

The Improved Reprocessing of the IRAS Survey (IRIS) images at 12, 25, 60 and 100~\micron{} show significant brightness and extension of the emission of the S233 region, which are comparable to the other regions of star formation within the entire S231-235 complex. Far-IR emission of the S233 region has roughly the same rounded morphology with a radius of $\simeq4\arcmin$ at all four IRAS bands, with the central peak coinciding with the position of IRAS 05351+3549.

\subsection{Physical parameters of dust} \label{phys_par_IR}

In order to estimate the physical parameters of the dust we used the method described in \citet{Cichowolski2001}. According to this method, the total integrated infrared luminosity of IRAS 05351+3549 is as follows (assuming the distance to the S233 region as 2.3~kpc):

\begin{center}
\begin{equation}
L_{\mathrm{IR}} = 1.58 F_{\mathrm{tot}}d^{2}_{\mathrm{kpc}} = 2.280\times10^{3}~\mathrm{L_{\odot}}
\label{Eq:luminosity}
\end{equation}
\end{center}
Where $F_{\mathrm{tot}} = 1.3(F_{12}+F_{25})+0.7(F_{25}+F_{60})+0.2(F_{60}+F_{100}) = 272.81~\mathrm{Jy}$ is the integrated source flux density; $F_{12}, F_{25}, F_{60}, F_{100}$ are the flux densities of the IRAS source at 12, 25, 60 and 100 $\micron$.

The calculation of the total mass of radiatively heated dust leads to the following results:
\begin{center}
\begin{equation}
\mathrm{M_{d}}=m_{\mathrm{n}}F_{60}d^{2}_{\mathrm{kpc}}(B^{2.5}_{\mathrm{n}}-1) = 0.7~\mathrm{M_{\bigodot}}
\label{Eq:mass_dust}
\end{equation}
\end{center}
In this formula $n$ is the emissivity index, related to the absorption efficiency of the dust ($k_{\mathrm{v}}\propto \nu^{n}$, normalized to dust absorption coefficient $40~\mathrm{cm^{2}g^{-1}}$ at 100~\micron). We used a value of $n=1.5$ which is typical for the \ion{H}{II} regions \citep{Rodon2010} and $m_{1.5}=0.3\times10^{-6}$. $F_{60}$ is the flux density of the IRAS sources at 60~\micron{} and $B_n$ is the modified Planck function, given by the equation $B_n=1.667^{3+n}(F_{100}/F_{60})=22.433$.

Using a dust-to-gas ratio of 0.01 \citep{Draine} we estimated the gas mass to be 70~\ms. However, this mass is not necessarily equal to the gas mass that in principle can be estimated from CO emission, since infrared emission may trace a different volume than that traced by CO emission.

The last step is a calculation of the dust temperature using the following formula:
\begin{center}
\begin{equation}
T_{\mathrm{d}}=\frac{95.94}{\ln{B_n}} = 30.8~\mathrm{K}
\label{Eq:mass_dust}
\end{equation}
\end{center}
This value is typical in star-forming regions \citep{Cappa08,Cichowolski09,Vasquez10} and consistent with the gas temperature in the nearby star-forming region S235 \citep{Kirsanova14} and S233-IR \citep{Schreyer96}.

\begin{figure}
\includegraphics[width=77mm]{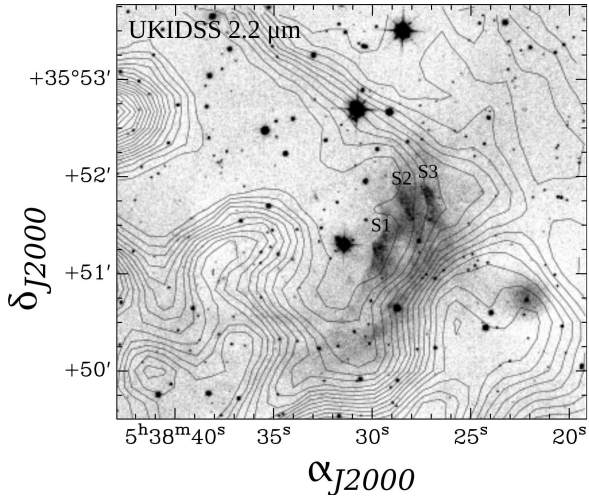}
\caption{Contours of the $^{12}$CO (2-1) emission of the S233 region with velocity $-$15.6~km~s$^{-1}$, overlaid on UKIDSS 2.2~\micron{} K-band image, showing interaction of the S233 \ion{H}{II} region with the ambient molecular cloud. Contour levels start from 1.5 K, stepping by 0.5 K. Infrared condensations S1, S2 and S3 are marked with labels.}
\label{co_border}
\end{figure}


\subsection{Ambient molecular cloud}

The S231-235 complex is a warm giant molecular cloud with brightness temperatures of the $^{12}$CO (2-1) line up to 50 K. Most of CO emission has main beam antenna temperatures $T_{\mathrm{mb}} \lesssim 15$~K. The complex was observed earlier in the $^{12}$CO (1-0) and $^{13}$CO (1-0) lines by \citet{Heyer96a}.

It appears that molecular gas in the S233 region \citep{Heyer96a} is the part of the large-scale filament that lies between the expanding S231 \ion{H}{II} region and the void at the west of S233. This void shows a circle-like morphology in maps given by \citet{Dame}.

A hint suggesting a physical interaction between the S233 \ion{H}{II} region and ambient molecular clouds is seen in the velocity range from $-$16.1 to $-$14.5 km~s$^ {-1}$.  The irregular border of CO emission at the west of the S233 region may be physically associated with the S233 \ion{H}{II} region (see Figure~\ref{co_border}). It is possible that the ionization front moves perpendicular to the observer in this velocity range. It should be noted that there is a large amount of material on the west of the S233 and there is a void at the east.

It is possible that the S1 peak and IRAS source are the result of the interaction between the \ion{H}{II} region and ambient molecular gas at a velocity of $\simeq$~$-16$~km~s$^{-1}$. There also exists a compact red-shifted component of gas at a velocity of $\simeq$~$-13$~km~s$^{-1}$, described in Section~\ref{molecular_clump}. The S1 peak also could be physically associated with this component. It may not be possible to draw any firm conclusion about whether the S1 IR peak is physically related to the molecular gas at a velocity of $-16$~km~s$^{-1}$ or to a redshifted component of gas at a velocity of $-13$~km~s$^{-1}$.

\begin{figure}
\includegraphics[width=80mm]{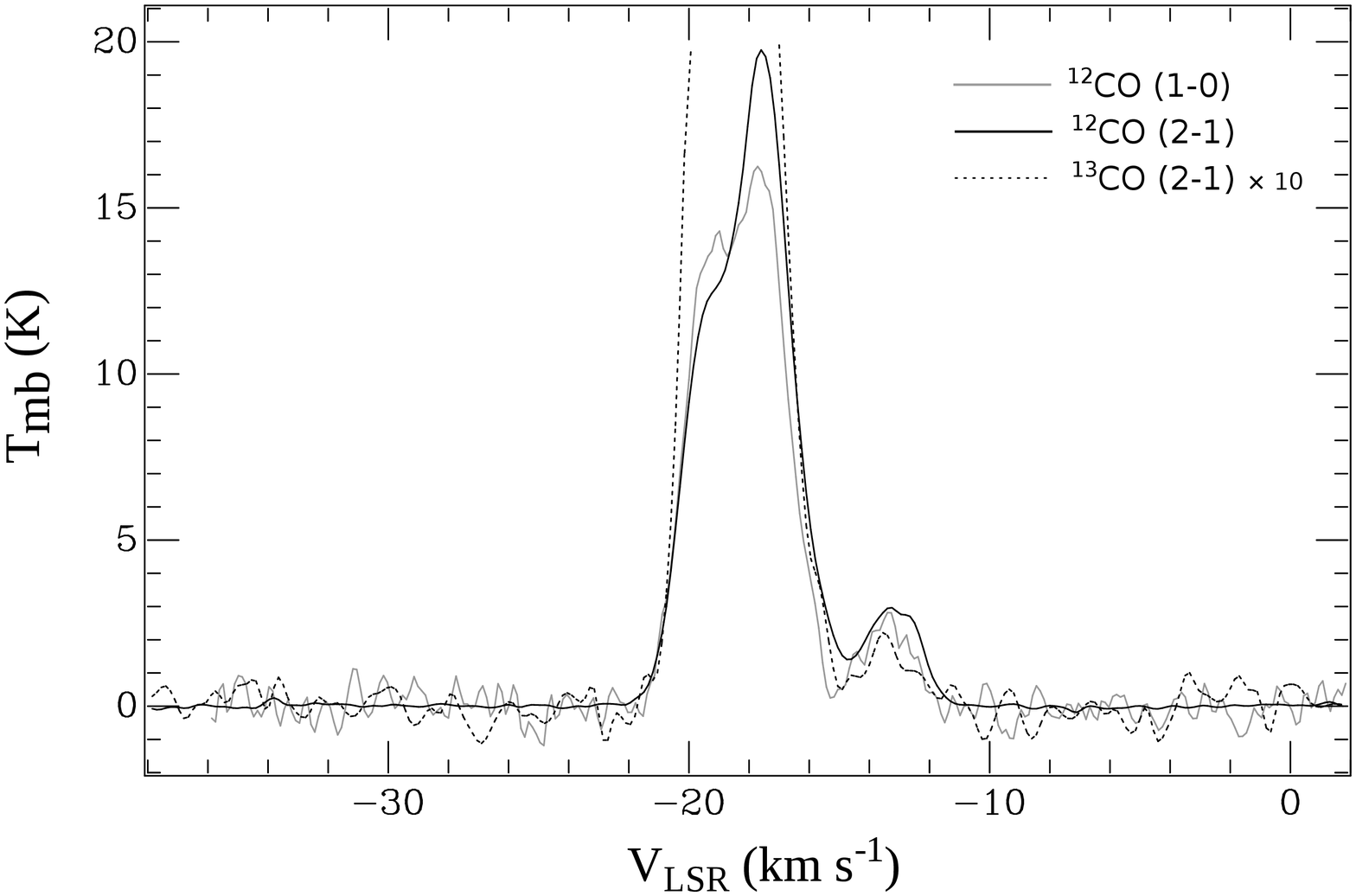}
\caption{Line profiles of CO in the direction of S233 ($\alpha_{\mathrm{J2000}}=05^{\rmn{h}}38^{\rmn{m}}28\fs 5~\delta_{\mathrm{J2000}}= +35\degr 50\arcmin 51\arcsec$). The black solid, gray solid and black dashed lines represent the $^{12}$CO (2-1), $^{12}$CO (1-0), and $^{13}$CO (2-1) lines. The data cubes are convolved to the same beam size of 47" for comparison. Hanning smoothing by 3 points was applied to the $^{12}$CO (1-0) line. The profile of $^{13}$CO (2-1) is multiplied by a factor of 10 to reveal low-brightness emission of the clump.}
\label{profile}
\end{figure}

The line intensity ratio of $^{12}$CO(2-1)/$^{13}$CO(2-1) in the S233 region ranges from 3 to 15, depending on the velocity shift from the line peak. Using the main beam antenna temperature of the $^{12}$CO line peak, we obtained a lower limit for the line ratio $^{12}\rm{CO}/^{13}\rm{CO} \simeq 3$. Using the values of antenna temperature away from the line peak, we got the higher line intensity ratios $^{12}\rm{CO}/^{13}\rm{CO} \simeq 10-15$. This is direct evidence that the $^{12}$CO line is optically thick.

\begin{figure*}
\begin{minipage}{140mm}
\includegraphics[width=140mm]{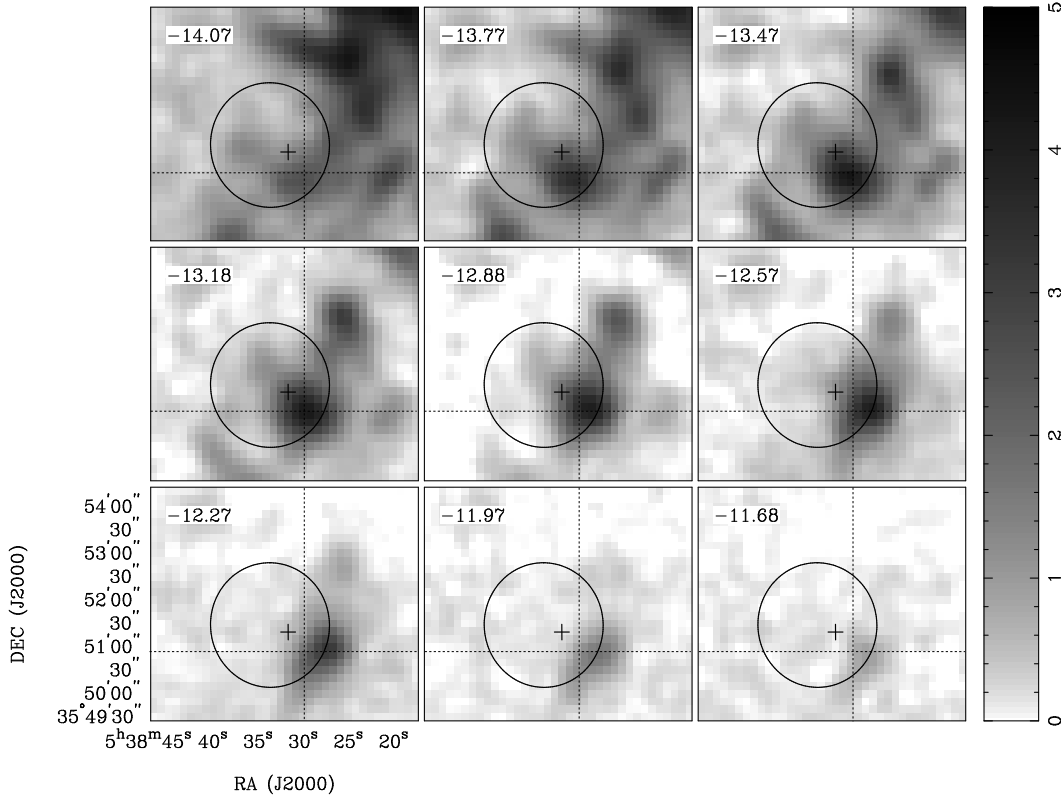}
\caption{ $^{12}$CO~(2-1) channel maps of the clump in the S233 region. The velocity of each panel is shown at the top-left corner. The ellipse represents the visual extent of the S233 \ion{H}{II} region. The cross indicates the position of the ionizing star. The vertical and horizontal lines exhibit the position of the intensity peak of the clump at $-$13.18 km~s$^{-1}$.}
\label{channel_map}
\end{minipage}
\end{figure*}
	
\subsection{Molecular clump in the S233 region} \label{molecular_clump}

Figure~\ref{profile} shows the profiles of the CO molecular line in the direction of the S233 region ($\alpha_{\mathrm{J2000}}=05^{\rmn{h}}38^{\rmn{m}}28\fs 5;~\delta_{\mathrm{J2000}}= +35\degr 50\arcmin 51\arcsec$). The primary component exhibits velocities in the range from $-$20 to $-$15 km~s$^{-1}$, and a weaker  red-shifted component is peaked at a velocity of $\sim$~$-13$~km~s$^{-1}$. The red-shifted component is pronounced in the $^{12}$CO~(2-1), $^{13}$CO~(2-1), and $^{12}$CO~(1-0) lines. Channel maps of the red-shifted component are presented in Figure~\ref{channel_map}.

\begin{figure}
\includegraphics[width=85mm]{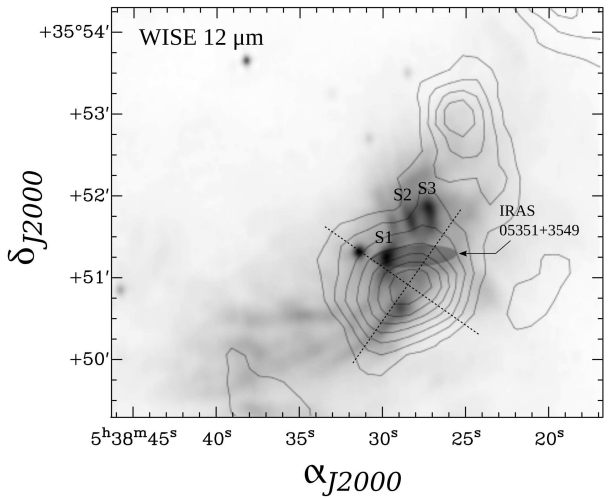}
\caption{Contours of the $^{12}$CO (2-1) emission, integrated from $-14.6$ to $-11.0$ km~s$^{-1}$, showing the molecular clump in the S233 region. The contour levels span from 2.22 K km~s$^{-1}$ to 6.66 K km~s$^{-1}$. The background is the WISE 12 \micron{} image. Filled ellipse shows uncertainty in position of IRAS source 05351+3549. Intensity profiles along the perpendicular dotted lines were used for the clump size determination (see Section~\ref{molecular_clump}).}
\label{cont}
\end{figure}

From the integrated image of the red-shifted component we estimate the spatial extent of this component using Gaussian fits to the spatial intensity profile along two perpendicular directions (see Figure~\ref{cont}). The source size is defined as FWHM of the fitted Gaussian functions. Derived values are $d_1~=~1\arcmin 11\arcsec \pm 4\arcsec$ and $d_2=1\arcmin 4\arcsec \pm 6.6\arcsec$, where the first intensity profile $d_1$ is obtained along the line passing through the ionizing star and the intensity peak of the red-shifted component. The second profile $d_2$ is obtained in the perpendicular direction. The source size appears to be approximately twice that of the telescope beam size $\theta=32\arcsec$, therefore the red-shifted component is not a point source. We conclude that this red-shifted component corresponds to a separate clump of gas. Assuming the distance to the S233 region $D=2.3\pm0.3$ kpc, the diameter of the clump is  $d_1=0.26\pm0.037$~pc, $d_2 = 0.23\pm0.038$~pc.

The main beam brightness temperatures of the clump emission are $3.42$~K, $2.79$~K, and $0.26$~K in the $^{12}$CO~(2-1), $^{12}$CO~(1-0), and $^{13}$CO~(2-1) lines, respectively. The clump emission shows a clear velocity gradient (see Figure~\ref{channel_map}). The minimum velocity at which the clump becomes visible in the $^{12}$CO (2-1) line is $-$14.3~km~s$^{-1}$, and the maximum one is $-$11.5~km~s$^{-1}$. The position of the intensity peak of the clump shifts by $\sim$ 1\arcmin{} when the radial velocity changes from from $-$14.3~km~s$^{-1}$ to $-$11.5~km~s$^{-1}$. At the distance of 2.3~kpc this velocity shift corresponds to a gradient of $\omega~\approx~1.8~\rm{km~s^{-1}~pc^{-1}}$.

The clump emission is located around the S1 IR condensation described in Section~\ref{mid_IR}. The existence of the gradient in the clump emission is probably a result of the interaction with the expanding \ion{H}{II} region. The position of the clump at the median velocity ($-$13~km~s$^ {-1} $) coincides with the position of the S1 mid-IR peak. As the velocity increases to $-$11~km~s$^{-1}$ the intensity peak of the clump shifts to the west relative to the S1 condensation and the edge of the \ion{H}{II} region. As the velocity decreases to $-$14.6 km~s$^{-1}$ the intensity peak of the clump shifts to the south relative to S1 (see Figure~\ref{channel_map}). These trends show that the clump has a complex morphology with the S1 mid-IR peak in the center.

\begin{table}
\caption{ Parameters of CO line profiles of the S233 molecular clump in direction $\alpha_{\mathrm{J2000}}=05^{\rmn{h}}38^{\rmn{m}}28\fs 5, \delta_{J2000}= +35\degr 50\arcmin 51\arcsec$. The first column is the observed main beam temperature of line, the second column is the central velocity of observed line and the third column is the FWHM of observed line. Standard error for each parameter is obtained from the Gaussian fit.}
\label{symbols}
\begin{tabular}{@{}lccccccc}
\hline
CO line & $T_{\mathrm{peak}}$ & $V_0$ & $\Delta V$ \\
& $(\mathrm{K})$ & $(\mathrm{km~s^{-1}})$ & $(\mathrm{km~s^{-1}})$ \\
\hline
$^{12}$CO (1-0) & 2.79 (0.6) & -13.48 (0.154) & 2.16 (0.39) \\
$^{12}$CO (2-1) & 3.42 (0.07) & -13.21 (0.015) & 2.20 (0.04) \\
$^{13}$CO (2-1) & 0.26 (0.04) & -13.36 (0.081) & 1.88 (0.20) \\
\hline
\end{tabular}
\label{tb1}
\end{table}

The clump is seen in three CO lines, but the emission is rather weak, so the noise levels prevent a detailed analysis of CO excitation using three lines simultaneously. So we applied only a simple analysis assuming LTE conditions and a uniform single gas layer. These assumptions allow us to estimate an optical depth using the main-beam brightness temperatures of the $^{12}$CO and $^{13}$CO in (2-1) lines (see Table \ref{tb1}) using the following equation:

\begin{center}
\begin{equation}
\frac{T_{12}}{T_{13}} = \frac{1-e^{-\tau}}{1-e^{-\tau / X}},
\label{Eq:ratio}
\end{equation}
\end{center}
here $X=60\pm10$ is the relative abundance of [$^ {12} $CO]/[$^{13}$CO] at the distance of 2.3 kpc \citep{Langer}. Using the observed value of $T_ {12} /T_{13} = 13.15\pm2.04$ and solving Equation (\ref{Eq:ratio}) numerically, we estimate that the optical depth of the $^{12}$CO~(2-1) line is in the range from 3.3 to 6.6. This shows that the $^{12}$CO line is optically thick.

For the extended sources the sum of the brightness temperatures of the optically thick line and background, $T_{12}+T_{\mathrm{bg}}$, provides an estimate of the gas kinetic temperature. In the case of the molecular clump both the $^{12}$CO~(2-1) and (1-0) lines have rather low values of main beam temperature $T_{\mathrm{mb}}\simeq3$~K. 
Correcting the observed radiation temperature for the Planck law, i.e., $T_\rmn{mb} = J_\rmn{\nu}(T) = (h \nu/k)/[\exp(h \nu/kT)-1]$, the derived excitation temperature for the $^{12}$CO (2-1) line is ~8 K for an observed peak line intensity of 3.4 K and assuming only the cosmic background of 2.73 K.  A value of 8 K is a lower limit to the gas kinetic temperature, because (1) the transition may be sub-thermally excited; (2) the continuum background may include some thermal dust continuum; and (3) the source may be only partially resolved so the observed brightness temperature is beam-diluted.  
The latter is in accord with recent observational results showing that the formation of low mass stars tends to occur in  filaments with a characteristic size of about 0.1 pc \citep{Arzoumanian} which is smaller than the linear size of the beam in the present observations. Observations with a higher spatial resolution and signal-to-noise ratio are necessary to reveal the situation.

Following \citet{MacLaren}, we estimated the virial mass of the clump using the linewidth of the $^{13}$CO line (1.88 km~s$^{-1}$) and the mean clump radius of 0.12 pc. Derived values range from 53 to 89~\ms, depending on the assumed density distribution as a function of the radial distance from the clump center $r$. The virial mass is calculated to be 53,~80~and~89~\ms{} in the cases of a constant density, density $\propto 1/r$ and $\propto 1/r^2$, respectively. All these values are of the same magnitude as the mass of the IRAS source 05351+3549, derived from IRAS flux densities (70~\ms, see Section~\ref{phys_par_IR} for details). This agreement indicates that the clump may be gravitationally bound if the IRAS source is physically associated with the CO clump.

\begin{figure}
\includegraphics[width=80mm]{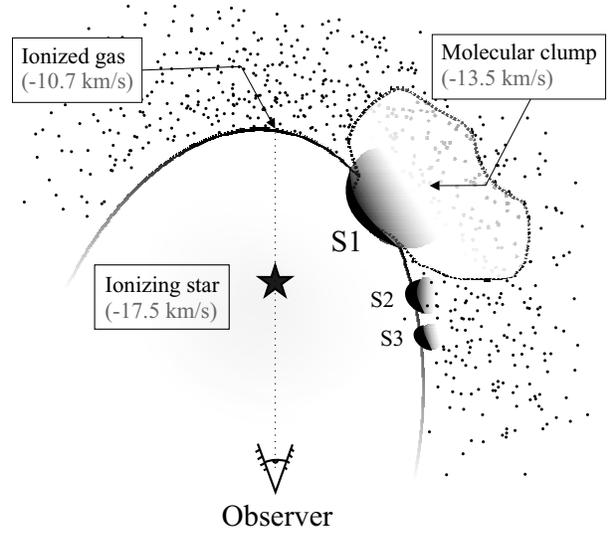}
\caption{Schematic representation of the S233 \ion{H}{II} region constructed from the UKIDSS image (Figure~\ref{ukidss}) and velocities of the ionizing star, ionized gas (Section~\ref{sec:rad_vel}) and molecular clump (Section~\ref{molecular_clump}). Dotted region shows molecular gas with the range of velocities from $-$20 to $-$15~km~s$^{-1}$. }
\label{clipart}
\end{figure}

\section{Discussion}
\subsection{Morphology of the S233 \ion{H}{II} region}

It is seen in the optical images (see Figure~\ref{ukidss}) that the S233 region has a cometary shape with the sharp edge situated in the south-west (see, e.g. \citet{Cohen}). This morphology could be due to the motion of the ionizing star relative to the ambient gas, or to inhomogeneity of the ambient cloud. In the Section \ref{sec:rad_vel} we have shown that the velocity of the ionizing star ($V_{\mathrm{LSR}}=-17.5\pm1.4$~km~s$^{-1}$) lies within the velocity range of the ambient molecular gas ($V_{\mathrm{LSR}}$ from $-20$ to $-15$ km~s$^{-1}$) that surrounds the star. So, it is more plausible that the central ionizing star was born in the molecular cloud, not that the star entered this cloud from the outside. Thus, the cometary morphology the S233 region seen in DSS Red image probably results from an inhomogeneous distribution of the ambient molecular gas (see Figure~\ref{clipart}). If the material to the south-west from the star is denser than the material to the north-east, then the photoionization and shock waves propagating to the south-west meet more resistance, thus producing the brighter part of the cometary-like nebula.

The condensations S1, S2 and S3 discussed in the Section \ref{mid_IR} might be formed by the expansion of the \ion{H}{II} region. In the vicinity of S1 the gas clump could be dense enough that its material cannot be ionized or pushed away by expansion of the S233 \ion{H}{II} region. The images clearly show that the condensations are oriented tangentially with respect to the direction towards the ionizing star. Thus, these condensations might be created by the photoionization and shock waves expanding from the S233 \ion{H}{II} region.

\subsection{A scenario for star formation in the S233 region}

In the S233 region we see evidence for interaction between the \ion{H}{II} and the ambient molecular cloud, which could trigger star formation. There is a compact infrared cometary-like condensation S1, which faces the ionizing star and exhibits emission at 24~\micron{} which comes primarily from heated dust. This condensation coincides in position with the IRAS source 05351+3549. This source displays flux density increasing with wavelength, with colours corresponding to those of a YSO (see Section~\ref{sec_IRAS}).  There is also a molecular clump in the direction of the S1 condensation, seen in $^{12}$CO emission red-shifted relative to that of the parent molecular cloud.

It is known that the expansion of an \ion{H}{II} region can trigger star formation. There are different scenarios of star formation triggered by the \ion{H}{II} region expansion. One of the commonly considered scenarios is the `collect-and-collapse' \citep{Elmegreen77}. In this scenario neutral gas in the layer between the ionization and shock fronts becomes gravitationally unstable which eventually leads to star formation. In the ideally symmetric case this neutral gas is expected to be observed as a ring structure around the \ion{H}{II} region. The young stellar clusters are expected to be located between the PDR and the parent molecular gas.

Following \citet{Kirsanova14} we consider the S233 \ion{H}{II} region within the framework of the theoretical one-dimensional model described in \citet{Whitworth94}. The model allows an estimate of the main characteristics of the `collect-and-collapse' process: starting time of the dense shell fragmentation (\tfrag), the radius of the shell (\rfrag) at this moment and the mass of the fragments (\mfrag). We also calculate the radius of the initial Str\"omgren sphere \rs~\citep{Dyson} and the \ion{H}{II} region age (using Eq.4 from \citet{Whitworth94}). However, the S233 region exhibits a non-uniform distribution of the ambient molecular material. Thus, the one-dimensional model with a homogeneous ambient density distribution cannot be considered as strictly applicable to this region.  Since there is no dense gas to the north-east of the \ion{H}{II}, the pressure in the \ion{H}{II} region is lower than in the model. Therefore we use the model of \citet{Whitworth94} to make only rough estimates.

\begin{table}
\caption{Characteristics of the one-dimensional model simulating the `collect-and-collapse’ process for the S233 \ion{H}{II} region. Parameters used for calculation of the triggering process are: $\log N_c = 47.0$, $R_\mathrm{shell}=0.67$~pc, $T_\mathrm{kin}=25$~K (see explanations in the text).}
\begin{tabular}{lccccc}
\hline
$n_0$, cm$^{-3}$ & $10^2$ & $5\cdot10^2$ & $10^3$& $3\cdot10^3$& $4\cdot10^4$ \\
\hline
\rs, pc & 0.74 & 0.25 & 0.15 & 0.05 & 0.01 \\
\rfrag, pc & 15.9 & 6.6 & 4.5 & 1.9 & 0.6 \\
\tfrag, Myr & 9.1 & 4.4 & 3.2 & 1.5 & 0.6 \\
\mfrag, \ms & 560 & 270 & 200 & 100 & 40 \\
Age, Myr & 0.03 & 0.06 & 0.09 & 0.2 & 0.6 \\
\hline
\end{tabular}
\label{tab:CC}
\end{table}

Input parameters of the one-dimensional model are the number of ionizing photons, the size of the \ion{H}{II} region, the kinetic temperature, and the gas density. The number of ionizing photons for the S233 ionizing star is $N_c = 10^{47.0}$ (see Section~\ref{sec:spec_analys}). For the calculations we used a fixed value of the \ion{H}{II} region radius. The angular diameter of the S233 region is 120\arcsec, taken from the optical measurements \citep{Sharpless}. Published distances to the objects in this complex vary from 1.6 to 2.3 kpc \citep{Reiputh}. Thus, the linear radius of the S233 region vary from 0.46~pc ($D=1.6$~kpc) to 0.67~pc ($D=2.3$~kpc). We adopt the value of 0.67~pc as a representative radius of the S233 \ion{H}{II} region. We adopt 25~K as a typical temperature in the S231-235 complex according to measurements of \citet{Kirsanova14, Schreyer96}. Note that the parameter estimates do not depend critically on the adopted temperature. For the gas density we consider several values from 100~cm$^{-3}$ to 40000~cm$^{-3}$. The age of the \ion{H}{II} region becomes comparable to \tfrag{} only with the highest assumed value of the gas density. Results of the calculations are given in Table~\ref{tab:CC}.

Analyzing the results of these calculations, we conclude that the `collect-and-collapse’ scenario in the S233 region could work if the average gas density in the region has the value $n_0~\simeq~4\cdot10^4$~cm$^{-3}$. With the density this high the fragmentation time becomes comparable to the \ion{H}{II} region age. The estimate of the mass of the clump in that case is \mfrag~$\approx40$~\ms, which is comparable with the gas mass estimate from the dust emission (70~\ms, see Section~\ref{phys_par_IR}). We cannot estimate the gas density in the S233 region from our data alone. However, the nearby S235~\ion{H}{II} region exhibits an average gas density of~$\approx$~500~cm$^{-3}$ only (see Table 1 in \citet{Kirsanova14}). Thus, the value of~$4\cdot10^4$~cm$^{-3}$ is probably too high for the average gas density in the S233~\ion{H}{II} region. Based on calculations using one-dimensional model of \citet{Whitworth94}, we conclude that the `collect-and-collapse' process is not likely to take place in the S233 region.

Another scenario for triggered star formation is contraction of a pre-existing clump due to radiation-driven implosion \citep{Lefloch94, Kessel-Deynet, Miao} or to a shock wave from the expanding \ion{H}{II} region \citep{Boss95}. This scenario predicts a random distribution of molecular clumps around the \ion{H}{II} region. The signature of this process is the presence of the bright rims surrounding cometary globules and condensations similar to those considered, e.g. by \citet{Thompson04}. Star formation could take place in these globules. In the case of the S233 \ion{H}{II} region we observe a bright rim with a cometary-like infrared condensation S1, facing the S233 ionizing star (see Figure~\ref{ukidss}).

\citet{Miao} conducted a three-dimensional hydrodynamical modeling to study the radiation-driven implosion effects of massive stars on surrounding molecular clouds. They showed that different morphologies of the bright rim represent different stages in the evolution of the cloud. The S1 condensation has a shape that is very similar to that of a molecular clump at stage I, according to the classification of evolutionary stages by \citet{Miao}. The distribution of infrared emission in the WISE wavelength bands shows that the gas in S1 is hotter on the side that faces the ionizing star (figure~\ref{wise}). At present the S1 condensation is elongated tangentially but it is possible that eventually condensation S1 will become more elongated in the radial direction and will become a classical cometary globule \citep{Lefloch94}.
The morphology of the optical and infrared emission and their spectral characteristics show that S1 represents a site of active interaction of the ionizing star emission with the dense gas clump. The S1 peak is associated with the IRAS point source, and probably is situated at the isolated molecular clump. It is possible, therefore, that star formation in the S1 region takes place because the shock wave from the \ion{H}{II} region has compressed the clump of  ambient material.

\section{Conclusions}

We have studied a star-forming region which includes an isolated B0.5~V ionizing star surrounded by the \ion{H}{II} region S233. The ionizing star has a sufficiently large Lyman continuum flux to account for the observed radio flux density of the S233 \ion{H}{II} region. The velocity of the ionizing star is determined to be $\rm{V_{\mathrm{LSR}}}=-17.5 \pm 1.4~\rm{km~s^{-1}}$. Comparing the velocity of the ionizing star with that of the ambient molecular gas (from $-$20 to $-$15 km~s$^{-1}$), we conclude that this star was formed from the material of the ambient gas cloud.

In the Section \ref{sec:rad_vel}, we have shown that the S233 \ion{H}{II} region expands with a velocity of $6.8\pm2.2$ km~s$^{-1}$ and interacts dynamically with the ambient molecular cloud material. We find evidence for interaction between S233 and the ambient molecular cloud, which could trigger star formation:

\begin{enumerate}

\item UKIDSS and WISE images reveal an extended region of infrared emission containing the compact condensation S1 (up to 24 \micron) and the bright-rimmed structure, perpendicular to the ionizing star. The bright-rimmed structures with infrared condensations are common in star formation triggered by radiatively-driven implosion \citep{Thompson04}. The dust component of the infrared condensation S1 reflects visible light (DSS Blue) from the ionizing star, which supports the idea of physical proximity of the star and the infrared source.

\item IRAS source 05351+3549 coincides in position with the compact infrared source S1 in the WISE and UKIDSS images. The flux of the IRAS source increases with wavelength (up to 414 Jy at 100 \micron) and the colours of the IRAS source correspond to those of a YSO in the colour-colour diagram given by \citet{Straizys07}. The dust temperature of the infrared source, $T_{\mathrm{d}}=30.8~\mathrm{K}$ is in good agreement with the gas temperature of star forming clumps in the nearby star-forming region S235 \citep{Kirsanova14}. The gas mass of the IRAS source is found to be 70 $\mathrm{M}_{\mathrm{\odot}}$.

\item CO study: The infrared condensation S1 coincides in position with the clump of molecular gas having a mean velocity \mbox{$-$13.1~km~s$^{-1}$}. The main gas component in the region has velocity in the interval from $-$20 km~s$^{-1}$ to $-15$~km~s$^{-1}$, so the clump may have been `disturbed' by the \ion{H}{II} region and is now moving away from us with a velocity which exceeds the velocity of `undisturbed' gas. The clump emission is observed at velocities from $-13.4~\mathrm{km~s^{-1}}$ to~$-11.8~\mathrm{km~s^{-1}}$ and shows a velocity gradient $\omega~\approx~1.8~\rm{km~s^{-1}~pc^{-1}}$. The optical depth of the clump in the $^{12}$CO lines is in the range $\tau~=~3.3 - 6.6$. The virial mass of the clump ranges from 53 to 89~\ms.
\end{enumerate}

We investigate the nature of the triggering mechanism for star formation in the S233 \ion{H}{II} region vicinity. Analysis of calculations using the one-dimensional model given in \citet{Whitworth94} indicates that the `collect-and-collapse' scenario in the S233 region could be realized if the average gas density in the region has the value $n_0~\simeq~4\cdot10^4$~cm$^{-3}$. This value is probably too high for the average gas density in the S233~\ion{H}{II} region. Thus, we conclude that the `collect-and-collapse' process is not likely to take place in the S233 region.

A more plausible scenario for triggering of star formation in the S233 \ion{H}{II} region is the `collapse of the pre-existing clump’. The signature of this process is the presence of the bright rims surrounding cometary globules and condensations. In the case of the S233 \ion{H}{II} region vicinity we observe a bright rim around the cometary-like condensation S1 seen in the infrared continuum emission. This rim is facing the S233 ionizing star (see Figure~\ref{ukidss}). Therefore, we suggest that star formation of the `collapse of the pre-existing clump’ type is taking place in this condensation.

\section{Acknowledgments}
The study has been supported by the Ministry of Education and Science of the Russian Federation within the framework of the research activities (project no. 3.1781.2014/K).

The Heinrich Hertz Submillimeter Telescope is operated by the Arizona Radio Observatory, a unit of Steward Observatory at The University of Arizona. This work was supported in part by U.S. National Science Foundation grant AST-1140030 to The University of Arizona.

We are very grateful to Chris Brunt for providing \mbox{CO(1-0)} FCRAO data. The Five College Radio Astronomy Observatory was supported by NSF grant AST 05-40852.

This publication makes use of data products from the Wide-field Infrared Survey Explorer \citep{Wright}, which is a joint project of the University of California, Los Angeles, and the Jet Propulsion Laboratory/California Institute of Technology, funded by the NASA.

This work also uses data products from UKIRT Infrared Deep Sky Survey (UKIDSS) archive. The UKIDSS project is defined in \citet{Lawrence}. UKIDSS uses the UKIRT Wide Field Camera \citep{Casali}. The photometric system is described in \citet{Hewett}, and the calibration is described in \citet{Hodgkin}. The pipeline processing and science archive are described in \citet{Hambly}.

\footnotesize{

}

\label{lastpage}


\begin{thebibliography}{99}
\bibitem[\protect\citeauthoryear{Arzoumanian et al.}{2011}]{Arzoumanian} Arzoumanian D., et al., 2011, A\&A, 529, L6
\bibitem[\protect\citeauthoryear{Becker \& Butler}{1990}]{Becker90} Becker S.~R., Butler K., 1990, A\&A, 235, 326 
\bibitem[\protect\citeauthoryear{Bertelli et al.}{2009}]{Bertelli} Bertelli G., Nasi E., Girardi L., Marigo P., 2009, A\&A, 508, 355 
\bibitem[\protect\citeauthoryear{Boss}{1995}]{Boss95} Boss A.~P., 1995, ApJ, 439, 224 
\bibitem[\protect\citeauthoryear{Cappa et al.}{2008}]{Cappa08} Cappa C., Niemela V.~S., Amor{\'{\i}}n R., Vasquez J., 2008, A\&A, 477, 173 
\bibitem[\protect\citeauthoryear{Cardelli, Clayton, \& Mathis}{1989}]{Cardelli} Cardelli J.~A., Clayton G.~C., Mathis J.~S., 1989, ApJ, 345, 245 
\bibitem[\protect\citeauthoryear{Casali et al.}{2007}]{Casali} Casali M., et al., 2007, A\&A, 467, 777 
\bibitem[\protect\citeauthoryear{Chan \& Fich}{1995}]{Chan} Chan G., Fich M., 1995, AJ, 109, 2611 
\bibitem[\protect\citeauthoryear{Cichowolski et al.}{2001}]{Cichowolski2001} Cichowolski S., Pineault S., Arnal E.~M., Testori J.~C., Goss W.~M., Cappa C.~E., 2001, AJ, 122, 1938 
\bibitem[\protect\citeauthoryear{Cichowolski et al.}{2009}]{Cichowolski09} Cichowolski S., Romero G.~A., Ortega M.~E., Cappa C.~E., Vasquez J., 2009, MNRAS, 394, 900 
\bibitem[\protect\citeauthoryear{Claret}{2004}]{Claret} Claret A., 2004, A\&A, 424, 919 
\bibitem[\protect\citeauthoryear{Cohen}{1980}]{Cohen} Cohen M., 1980, AJ, 85, 29 
\bibitem[\protect\citeauthoryear{Crowther}{1997}]{Crowther} Crowther P.~A., 1998, in Bedding T.~R., Booth A.~J., Davis J., eds, Proc. IAU Symp. no 189, Fundamental Stellar Properties: the Interactions between Observations and Theory. Kluwer, Dordrecht, p. 137
\bibitem[\protect\citeauthoryear{Dale et al.}{2009}]{Dale09} Dale J.~E., W{\"u}nsch R., Whitworth A., Palou{\v s} J., 2009, MNRAS, 398, 1537
\bibitem[\protect\citeauthoryear{Dame, Hartmann, \& Thaddeus}{2001}]{Dame} Dame T.~M., Hartmann D., Thaddeus P., 2001, ApJ, 547, 792 
\bibitem[\protect\citeauthoryear{Deharveng et al.}{2003}]{Deharveng03} Deharveng L., Lefloch B., Zavagno A., Caplan J., Whitworth A.~P., Nadeau D., Mart{\'{\i}}n S., 2003, A\&A, 408, L25
\bibitem[\protect\citeauthoryear{Deharveng et al.}{2008}]{Deharveng08} Deharveng L., Lefloch B., Kurtz S., Nadeau D., Pomar{\`e}s M., Caplan J., Zavagno A., 2008, A\&A, 482, 585
\bibitem[\protect\citeauthoryear{Dobashi et al.}{2005}]{Dobashi} Dobashi K., Uehara H., Kandori R., Sakurai T., Kaiden M., Umemoto T., Sato F., 2005, PASJ, 57, 1 
\bibitem[\protect\citeauthoryear{Draine et al.}{2007}]{Draine} Draine B.~T., et al., 2007, ApJ, 663, 866 
\bibitem[\protect\citeauthoryear{Dyson \& Williams}{1997}]{Dyson} Dyson J.~E., Williams D.~A., 1997, in Dyson J. E., Williams D. A., eds, The Physics of the Interstellar Medium. 2nd edn. The Graduate Series in Astronomy. Institute of Physics Publishing, Bristol
\bibitem[\protect\citeauthoryear{Elmegreen \& Lada}{1977}]{Elmegreen77} Elmegreen B.~G., Lada C.~J., 1977, ApJ, 214, 725 
\bibitem[\protect\citeauthoryear{Elmegreen}{1998}]{Elmegreen98} Elmegreen B.~G., 1998, ASPC, 148, 150
\bibitem[\protect\citeauthoryear{Galazutdinov}{1992}]{Galazutdinov} Galazutdinov, G.~A., 1992, SAO Preprint, 92
\bibitem[\protect\citeauthoryear{Gooch}{1995}]{Gooch} Gooch R., 1995, ASPC, 77, 144 
\bibitem[\protect\citeauthoryear{Grevesse \& Sauval}{1998}]{Grevesse} Grevesse N., Sauval A.~J., 1998, Space Sci. Rev., 85, 161 
\bibitem[\protect\citeauthoryear{Hambly et al.}{2008}]{Hambly} Hambly N.~C., et al., 2008, MNRAS, 384, 637 
\bibitem[\protect\citeauthoryear{Hewett et al.}{2006}]{Hewett} Hewett P.~C., Warren S.~J., Leggett S.~K., Hodgkin S.~T., 2006, MNRAS, 367, 454 
\bibitem[\protect\citeauthoryear{Heyer et al.}{1996}]{Heyer96a} Heyer M.~H., Carpenter J.~M., Ladd E.~F., 1996, ApJ, 463, 630 
 \bibitem[\protect\citeauthoryear{Heyer et al.}{1996}]{Heyer96b} Heyer M.~H., et al., 1996, ApJ, 464, L175
\bibitem[\protect\citeauthoryear{Hodgkin et al.}{2009}]{Hodgkin} Hodgkin S.~T., Irwin M.~J., Hewett P.~C., Warren S.~J., 2009, MNRAS, 394, 675 
\bibitem[\protect\citeauthoryear{Hubeny \& Lanz}{1992}]{Hubeny} Hubeny I., Lanz T., 1992, A\&A, 262, 501 
\bibitem[\protect\citeauthoryear{Hunter \& Massey}{1990}]{Hunter} Hunter D.~A., Massey P., 1990, AJ, 99, 846
\bibitem[\protect\citeauthoryear{Kang et al.}{2012}]{Kang} Kang J.-h., Koo B.-C., Salter C., 2012, AJ, 143, 75 
\bibitem[\protect\citeauthoryear{Kessel-Deynet \& Burkert}{2003}]{Kessel-Deynet} Kessel-Deynet O., Burkert A., 2003, MNRAS, 338, 545
\bibitem[\protect\citeauthoryear{Kilian}{1992}]{Kilian92} Kilian J., 1992, A\&A, 262, 171 
\bibitem[\protect\citeauthoryear{Kirsanova et al.}{2008}]{Kirsanova08} Kirsanova M.~S., Sobolev A.~M., Thomasson M., Wiebe D.~S., Johansson L.~E.~B., Seleznev A.~F., 2008, MNRAS, 388, 729 
\bibitem[\protect\citeauthoryear{Kirsanova, Wiebe, \& Sobolev}{2009}]{Kirsanova09} Kirsanova M.~S., Wiebe D.~S., Sobolev A.~M., 2009, Astron. Rep., 53, 611 
\bibitem[\protect\citeauthoryear{Kirsanova et al.}{2014}]{Kirsanova14} Kirsanova M.~S., Wiebe D.~S., Sobolev A.~M., Henkel C., Tsivilev A.~P., 2014, MNRAS, 437, 1593 
\bibitem[\protect\citeauthoryear{Langer \& Penzias}{1990}]{Langer} Langer W.~D., Penzias A.~A., 1990, ApJ, 357, 477 
\bibitem[\protect\citeauthoryear{Lanz \& Hubeny}{2007}]{Lanz} Lanz T., Hubeny I., 2007, ApJS, 169, 83 
\bibitem[\protect\citeauthoryear{Lawrence et al.}{2007}]{Lawrence} Lawrence A., et al., 2007, MNRAS, 379, 1599 
\bibitem[\protect\citeauthoryear{Lefloch \& Lazareff}{1994}]{Lefloch94} Lefloch B., Lazareff B., 1994, A\&A, 289, 559 
\bibitem[\protect\citeauthoryear{MacLaren et al.}{1988}]{MacLaren} MacLaren I., Richardson K.~M., Wolfendale A.~W., 1988, ApJ, 333, 821
\bibitem[\protect\citeauthoryear{Martins et al.}{2005}]{Martins05} Martins F., Schaerer D., Hillier D.~J., 2005, A\&A, 436, 1049
\bibitem[\protect\citeauthoryear{Masci \& Fowler}{2009}]{Masci} Masci F.~J., Fowler J.~W., 2009, ASPC, 411, 67 
\bibitem[\protect\citeauthoryear{Miao et al.}{2006}]{Miao} Miao J., White G.~J., Nelson R., Thompson M., Morgan L., 2006, MNRAS, 369, 143 
\bibitem[\protect\citeauthoryear{Oey et al.}{2005}]{Oey05} Oey M.~S., Watson A.~M., Kern K., Walth G.~L., 2005, AJ, 129, 393 
\bibitem[\protect\citeauthoryear{Panchuk et al.}{2009}]{Panchuk} Panchuk V.~E., Klochkova V.~G., Yushkin M.~V., Naidenov I.~D., 2009, J Opt Tech, 76, 2 
\bibitem[\protect\citeauthoryear{Parimucha \& Skoda}{2007}]{Parimucha} Parimucha, {\v S}., \& {\v S}koda, P.\ 2007, IAU Symposium, 240, 486
\bibitem[\protect\citeauthoryear{Pineda et al.}{2010}]{Pineda} Pineda A., et al, 2010, AJ, 721, 686 
\bibitem[\protect\citeauthoryear{Piskunov et al.}{1995}]{Piskunov} Piskunov N.~E., Kupka F., Ryabchikova T.~A., Weiss W.~W., Jeffery C.~S., 1995, A\&AS, 112, 525 
\bibitem[\protect\citeauthoryear{Reipurth \& Yan}{2008}]{Reiputh} Reipurth B., Yan C.-H., 2008, in ASP Conf. Ser. 402, San Francisco, CA: ASP, 869 
\bibitem[\protect\citeauthoryear{Rodon et al.}{2010}]{Rodon2010} Rodon J.~A., et al., 2010, A\&A, 518, L80 
\bibitem[\protect\citeauthoryear{Sault, Teuben, \& Wright}{1995}]{Sault} Sault R.~J., Teuben P.~J., Wright M.~C.~H., 1995, ASPC, 77, 433 
\bibitem[\protect\citeauthoryear{Schneider et al.}{2010}]{Schneider10} Schneider N., et al., 2010, A\&A, 518, L83 
\bibitem[\protect\citeauthoryear{Schreyer et al.}{1996}]{Schreyer96} Schreyer K., Henning T., Koempe C., Harjunpaeae P., 1996, A\&A, 306, 267
\bibitem[\protect\citeauthoryear{Sharpless}{1959}]{Sharpless} Sharpless S., 1959, ApJS, 4, 257
\bibitem[\protect\citeauthoryear{Sim{\'o}n-D{\'{\i}}az}{2010}]{Simon-Diaz10} Sim{\'o}n-D{\'{\i}}az S., 2010, A\&A, 510, A22 
\bibitem[\protect\citeauthoryear{Snider et al.}{2009}]{Snider09} Snider K.~D., Hester J.~J., Desch S.~J., Healy K.~R., Bally J., 2009, ApJ, 700, 506 
\bibitem[\protect\citeauthoryear{Strai{\v z}ys \& Laugalys}{2007}]{Straizys07} Strai{\v z}ys V., Laugalys V., 2007, BaltA, 16, 327
\bibitem[\protect\citeauthoryear{Thompson, Urquhart, \& White}{2004}]{Thompson04} Thompson M.~A., Urquhart J.~S., White G.~J., 2004, A\&A, 415, 627
\bibitem[\protect\citeauthoryear{Urquhart et al.}{2004}]{Urquhart04} Urquhart J.~S., Thompson M.~A., Morgan L.~K., White G.~J., 2004, A\&A, 428, 723
\bibitem[\protect\citeauthoryear{Vasquez et al.}{2010}]{Vasquez10} Vasquez J., Cappa C.~E., Pineault S., Duronea N.~U., 2010, MNRAS, 405, 1976 
\bibitem[\protect\citeauthoryear{Walborn \& Fitzpatrick}{1990}]{Walborn} Walborn N., Fitzpatrick E.~L., 1990, PASP, 102, 1094 
\bibitem[\protect\citeauthoryear{Walker \& Cohen}{1988}]{Walker88} Walker H.~J., Cohen M., 1988, AJ, 95, 1801
\bibitem[\protect\citeauthoryear{Weikard et al.}{1996}]{Weikard96} Weikard H., Wouterloot J.~G.~A., Castets A., Winnewisser G., Sugitani K., 1996, A\&A, 309, 581
\bibitem[\protect\citeauthoryear{Whitworth et al.}{1994}]{Whitworth94} Whitworth A.~P., Bhattal A.~S., Chapman S.~J., Disney M.~J., Turner J.~A., 1994, MNRAS, 268, 291
\bibitem[\protect\citeauthoryear{Wouterloot \& Brand}{1989}]{Wouterloot} Wouterloot J.~G.~A., Brand J., 1989, A\&AS, 80, 149
\bibitem[\protect\citeauthoryear{Wright et al.}{2010}]{Wright} Wright E.~L., et al., 2010, AJ, 140, 1868 
\bibitem[\protect\citeauthoryear{Yushkin et al.}{2005}]{Yushkin} Yushkin M.V., Klochkova V. G., 2005, Preprint Spets. Astrofiz. Observ., 206 
\bibitem[\protect\citeauthoryear{Zavagno et al.}{2006}]{Zavagno06} Zavagno A., Deharveng L., Comer{\'o}n F., Brand J., Massi F., Caplan J., Russeil D., 2006, A\&A, 446, 171
\bibitem[\protect\citeauthoryear{Zavagno et al.}{2010}]{Zavagno10} Zavagno A., et al., 2010, A\&A, 518, L81 
\end{thebibliography}
\end{document}